\documentclass[11pt,a4paper]{article}
\title{\bf Heat kernel upper bounds under the generalized curvature(-dimension) inequality}
\author{Huai Qian LI \footnote{Email: huaiqianlee@gmail.com}
\vspace{3mm}\\
{\footnotesize Key Lab of Random Complex Structures and Data
Science, Academy of Mathematics and}\\
{\footnotesize  Systems Science, Chinese Academy of Sciences,
Beijing 100190, China} }
\date{}
\usepackage{amssymb,amsmath,amsfonts,amsthm,color,mathrsfs}

\def\R{\mathbb{R}}

\def\d{\textup{d}}

\def\supp{\textup{supp}}
\def\<{\langle}
\def\>{\rangle}
\def\Proof.{\noindent{\bf Proof. }}

\def\fin{\hfill$\square$}

\def\newdot{{\kern.8pt\cdot\kern.8pt}}

\newtheorem{theorem}{Theorem}[section]
\newtheorem{lemma}[theorem]{Lemma}
\newtheorem{corollary}[theorem]{Corollary}
\newtheorem{proposition}[theorem]{Proposition}

\newtheorem{definition}[theorem]{Definition}
\theoremstyle{definition}\newtheorem{remark}[theorem]{Remark}

\begin{document}

\maketitle
\makeatletter 
\renewcommand\theequation{\thesection.\arabic{equation}}
\@addtoreset{equation}{section}
\makeatother 

\begin{abstract}
In the sub-Riemannian manifolds, on the one hand, following Baudoin-Garofalo \cite{BaudoinGarofalo}, the upper bound for heat kernels associated to a class of locally subelliptic operators are given under the generalized curvature-dimension inequality with a negative curvature parameter; on the other hand, the argument combining Grigor'yan's integrated maximum principle with Wang's dimension-free Harnack inequality is also shown to derive the upper bound for the heat kernel under the generalized curvature inequality.
\end{abstract}



\section{Introduction}
The notion of Ricci curvature lower bounds in metric measure spaces using the mass transportation theory was proposed independently by Sturm \cite{Sturm1, Sturm2} and
by Lott--Villani \cite{LottVillani}. Both definitions are based on the convexity inequalities for functionals in the space of probability measures. But Juillet \cite{Juillet} pointed out that the theory developed in the above mentioned works seems not suitable for sub-Riemannian manifolds.


In 1983, Bakry--Emery \cite{BakryEmery} extend the notion of the Ricci curvature lower bound on the Riemannian manifold to the well-known curvature-dimension inequality with respect to (w.r.t. for short) diffusion operators, denoted by $CD(K,N)$ with $K\in\R$ and $N\in (0,\infty]$.  Recently, in the context of sub-Riemannian manifolds, Baudoin-Garofalo \cite{BaudoinGarofalo} introduced an interesting notion of generalized curvature-dimension inequality $CD(\rho_1,\rho_2,\kappa,d)$ (see \eqref{gcdi} below) w.r.t. a class of subelliptic operators, which is a generalization of the well-known Bakry--Emery curvature-dimension inequality. In that framework, the celebrated works by Yau \cite{Yau1, Yau2} and Li--Yau \cite{LiYau} on Riemannian manifolds with Ricci curvature bounded from below are generalized. Especially, the parabolic Harnack inequality and off-diagonal Gaussian upper bounds are given in the nonnegative curvature case, i.e., $\rho_1\geq 0$. See \cite{BaudoinBonnefont2,BaudoinGarofalo,BaudoinGordinaMelcher,BaudoinKim} for other applications. When $d=\infty$, the generalized curvature-dimension inequality becomes the generalized curvature inequality $CD(\rho_1,\rho_2,\kappa,\infty)$ (see \eqref{gci} below) which is a generalization of the Bakry--Emery curvature inequality $CD(K,\infty)$. In \cite{BaudoinBonnefont1}, under $CD(\rho_1,\rho_2,\kappa,\infty)$, the Poincar\'{e} inequality for the associated Dirichlet form, the log-Sobolev inequality and Wang's dimension-free Harnack inequality for the associated diffusion semigroup, and the HWI and isoperimetric  inequalities are studied.

The purpose of the present work is to give heat kernel upper bounds under the generalized curvature-dimension and curvature inequalities for a class of subelliptic diffusion operators on smooth sub-Riemannian manifolds, respectively. In Section 2, following closely the framework of Baudoin--Garofalo \cite{BaudoinGarofalo}, some preliminaries are given, including the generalized curvature-dimension inequality. In Section 3, the heat kernel upper bound is derived from the Li-Yau type Harnack inequality, under the generalized curvature-dimension inequality with a negative curvature parameter, i.e., $\rho_1<0$.  Finally, in Section 4, Grigor'yan's integrated maximum principle and Wang's dimension-free Harnack inequality are unified to derive the upper bound for the heat kernel under generalized curvature inequality with $\rho_1\in \R$.

\section{Preparations}
Let $M$ be a smooth connected finite dimensional manifold endowed with a smooth measure $\mu$ and let $L$ be a smooth second order diffusion operator on $M$ with real coefficients. We assume that $L$ is locally subelliptic (see \cite{FeffermanPhong} and \cite{JerisonS} for the definition and some properties of such operators) such that $L1=0$ and
$$\int_MfLg\,\d\mu=\int_M gLf\,\d\mu, \quad \int_MfLf\,\d\mu\leq 0,$$
for any $f,g\in C_c^\infty(M)$, the space of compactly supported infinitely differentiable functions on $M$.

Following Bakry--Emery \cite{BakryEmery}, define the {\it carr\'{e} du champ} of $L$, i.e., the first order symmetric quadratic differential operator, by
$$\Gamma(f,g)=\frac{1}{2}\{L(fg)-fL g-gL f\},\quad\mbox{for any }f,g\in C^\infty(M).$$
Here and in the sequel, we write $\Gamma(f)=\Gamma(f,f)$ for short. It is well known that there exists a canonical distance associated to $L$, defined by
\begin{eqnarray*}
d(x, y) = \sup\{|f(x)-f(y)|:\, f \in C^\infty(M),\, \Gamma(f)\leq 1\},\quad x, y \in M.
\end{eqnarray*}

There exists also an intrinsic distance associated to $L$ that can be defined via the notion of subunit curves (see \cite{FeffermanPhong}). An absolutely continuous curve $\gamma: [0,T]\rightarrow M$ is said to be subunit for the operator $L$ if, for every smooth function $f\in C^\infty(M)$,
$$\left|\frac{\d}{\d t}f(\gamma(t))\right|\leq \sqrt{\Gamma(f)(\gamma(t))}.$$
Then the subunit length of $\gamma$ is defined as $l_s(\gamma)=T$. Given $x,y\in M$, denote
$$S(x,y)=\{\gamma: [0,T]\rightarrow M: \gamma\,\ \hbox{is subunit for}\,\ L, \gamma(0)=x, \gamma(T)=y\}.$$
In this work, we assume that
$$S(x,y)\neq \emptyset,\,\ {\rm for\,\ any}\,\ x,y\in M.$$
Under this assumption, it is easy to prove that for $x,y\in M$,
$$d_S(x,y):=\inf\{l_s(\gamma): \gamma\in S(x,y)\}$$
defines a true distance on $M$. Furthermore, it is known from Lemma (5.29) in \cite{CarlenKusuokaStroock} that
$$d(x, y) = d_S(x, y), \quad x, y\in M,$$
and hence we can use indifferently either one of the distances $d$ or $d_S$.

Throughout this work, we assume that the metric space $(M,d)$ is complete.

In addition to the differential operator $\Gamma$, we assume that $M$ is endowed with another smooth symmetric bilinear differential operator,
denoted by $\Gamma^Z$, which satisfies that for $f,g,h\in C^\infty(M)$,
$$\Gamma^Z(fg, h) = f\Gamma^Z(g, h) + g\Gamma^Z(f, h),$$
and $\Gamma^Z(f):=\Gamma^Z(f, f)\geq 0$.

We now introduce the general assumptions that will be in force throughout the paper.\\
(H.1) There exists an increasing sequence $h_k\in C^\infty_c(M)$ such that $h_k\uparrow 1$ on $M$, and
$$\|\Gamma(h_k)\|_\infty + \|\Gamma^Z(h_k)\|_\infty\rightarrow 0,\,\ \hbox{as}\,\ k\rightarrow\infty,$$
where $\|\cdot\|_\infty$ is the essential supremum norm.\\
(H.2) For any $f\in C^\infty(M)$,
$$\Gamma(f, \Gamma^Z(f)) = \Gamma^Z(f, \Gamma(f)).$$

From now on, we denote the semigroup associated to $L$ by $\{P_t\}_{t\geq 0}$.\\
(H.3) The semigroup $P_t$ is stochastically complete, i.e., for $t\geq 0$, $P_t1=1$, and for every
$f\in C^\infty_c(M)$ and $T\geq 0$, it holds
$$\sup_{0\leq t\leq T}\{\|\Gamma(P_tf)\|_\infty  + \|\Gamma^Z(P_tf)\|_\infty  \}<\infty.$$

We define similarly as in the Riemannian case the following second order differential operators: for any $f,g\in C^\infty(M)$,
\begin{eqnarray*}
\Gamma_2(f,g)&=&\frac{1}{2}\left\{L\Gamma(f, g)-\Gamma(f,L g)-\Gamma(g,L f)\right\},\\
\Gamma_2^Z(f,g)&=&\frac{1}{2}\{L\Gamma^Z(f,g)-\Gamma^Z(f,L g)-\Gamma^Z(g, L f)\}.
\end{eqnarray*}
Likewise, set $\Gamma_2(f)=\Gamma_2(f,f)$ and $\Gamma_2^Z(f)=\Gamma_2^Z(f,f)$.

The final assumption is the following generalized curvature-dimension inequality introduced in \cite{BaudoinGarofalo}.
\begin{definition}\label{def}
We say that $M$ satisfies the generalized curvature-dimension
inequality $CD(\rho_1, \rho_2, \kappa, d)$ with respect to $L$ and $\Gamma^Z$ if there exist constants $\rho_1\in\R$, $\rho_2 > 0$, $\kappa \geq 0$
and $0 < d \leq \infty$ such that the inequality
\begin{equation}\label{gcdi}
\Gamma_2(f) + \nu\Gamma^Z_2(f)\geq \frac{(L f)^2}{d} + \left(\rho_1-\frac{\kappa}{\nu}\right)\Gamma(f) + \rho_2 \Gamma^Z(f)
\end{equation}
holds for every $f \in C^\infty(M)$ and every $\nu > 0$.
\end{definition}
\begin{remark}
It is understood in the previous definition that $CD(\rho_1, \rho_2, \kappa, \infty)$ means
\begin{equation}\label{gci}
\Gamma_2(f) + \nu\Gamma^Z_2(f)\geq \left(\rho_1-\frac{\kappa}{\nu}\right)\Gamma(f) + \rho_2 \Gamma^Z(f),
\end{equation}
which is called the generalized curvature inequality (see \cite{BaudoinBonnefont1}).
\end{remark}

When $\Gamma^Z=0$ and $\kappa=0$, \eqref{gcdi} and \eqref{gci} are the well-known Bakry--Emery curvature-dimension inequality $CD(\rho_1,d)$ and curvature inequality $CD(\rho_1,\infty)$, respectively. So, to some extent, the parameter $\rho_1$ plays the role of a lower bound of a sub-Riemannian generalization of the Ricci curvature (see the examples in \cite{BaudoinGarofalo}).

Let $L^p(M)=L^p(M,\mu)$ for $1\leq p\leq \infty$. Under the above assumptions, $L$ is nonpositive symmetric with respect to $\mu$ on $C_c^\infty(M)$. Following the arguments of  Strichartz \cite{Strichartz}, $L$ is essentially self-adjoint on $C_c^\infty(M)$, whose unique self-adjoint extension, also denoted by $L$, is the generator of the strongly continuous contraction semigroup $\{P_t\}_{t\geq 0}$ on $L^2(M)$, and, moreover,
\begin{equation}\label{contraction}
\|P_tf\|_{L^p(M)}\leq \|f\|_{L^p(M)},\quad 1\leq p\leq \infty.
\end{equation}
By H\"{o}mander's theorem (see e.g. \cite{Homander}), the map $(t,x)\mapsto P_tf(x)$ is smooth on $(0,\infty)\times M$ and for every $x\in M$,
$$P_tf(x)=\int_M p(t,x,y)f(y)\,\d\mu(y),\quad f\in C_c^\infty(M),$$
where $p(t,x,y)>0$ is the heat kernel associated to $P_t$, and such function is smooth in $(0,\infty)\times M\times M$ and symmetric in the spatial variables, i.e., $p(t,x,y)=p(t,y,x)$.

Throughout this work, let $B(x,r):=\{y\in M: d(y,x)<r\}$ be the metric ball in $(M,d)$ centered at $x\in M$ with radius $r>0$, and set
$$D=d\left(1+\frac{3\kappa}{2\rho_2}\right),\quad \rho^-_1=\max\{-\rho_1, 0\},$$
where $d, \rho_2, \kappa$ are parameters in Definition \ref{def}.

We should mention that a large class of sub-Riemannian manifolds with transverse symmetries satisfy the generalized curvature-dimension inequality \eqref{gcdi}, including the Sasakian manifold with horizontal Webster-Tanaka-Ricci curvature bounded from below, all Carnot groups with step two, and a class of principal bundles over Riemannian manifolds with the Ricci curvature bounded from below (see the examples in \cite{BaudoinGarofalo}). Recently, Wang \cite{Wany3} proposed an extension of the generalized curvature-dimension inequality, and generalized and partially improved the corresponding results in \cite{BaudoinBonnefont1}.

\section{Upper bounds for $p(t,x,y)$ under $CD(\rho_1,\rho_2,\kappa, d)$}
The off-diagonal Gaussian upper bounds for the heat kernel $ p(t,x,y)$ is given in \cite{BaudoinGarofalo} under the generalized curvature-dimension inequality $CD(\rho_1,\rho_2,\kappa, d)$ with $\rho_1\geq 0$ by the method of Li--Yau \cite{LiYau}. In this section, we give the upper bounds for $ p(t,x,y)$ in the case $\rho_1<0$,  following closely the aforementioned work.

At first, we generalize the Harnack inequality established in \cite{LiYau} for positive solutions to the heat equation $Lu-u_t=0$ on $M$, which are of the form $u(t,x)=P_tf(x)$ for some positive $f\in C_\infty^b(M):=C^\infty(M)\cap L^\infty(M)$ under the generalized curvature-dimension inequality \eqref{gcdi} with $\rho_1<0$ (see Theorem 7.1 in \cite{BaudoinGarofalo} for the case where $\rho_1\geq 0$). Before doing this, we need the following Li--Yau type inequality which is a generalization of the Li--Yau inequality in \cite{LiYau}, whose proof can be found in \cite{BaudoinGarofalo}.
\begin{lemma}\label{gradient}
Assume {\rm(H.1)}, {\rm(H.2)}, {\rm(H.3)} and \eqref{gcdi} holds with $\rho_1<0$. Let $f\in C^\infty_c(M)$ with $f > 0$, then for $t>0$, it holds
\begin{eqnarray*}
\Gamma(\log P_tf)+\frac{2\rho_2}{3}t\Gamma^Z(\log P_tf)\leq \left(\frac{D}{d}+\frac{2\rho_1^-}{3}t\right)\frac{LP_tf}{P_tf}+\frac{d(\rho_1^-)^2}{6}t+\frac{D\rho_1^-}{2}+\frac{D^2}{2dt}.
\end{eqnarray*}
\end{lemma}

From this lemma, we can immediately derive the following Li--Yau type Harnack inequality for the semigroup $P_t$.
\begin{theorem}\label{harnack}
Assume {\rm(H.1)}, {\rm(H.2)}, {\rm(H.3)} and \eqref{gcdi} holds with $\rho_1<0$. Given $(s,x), (t,y)$ $\in (0,\infty)\times M$ with $s<t$, for any $f\in C_\infty^b(M)$ such that $f>0$, it holds
\begin{eqnarray*}
(P_sf)(x)&\leq& (P_tf)(y)\left(\frac{t}{s}\right)^{\frac{D}{2}} \exp\left\{\frac{d(x,y)^2}{4(t-s)}\left(\frac{D}{d}+\frac{2\rho_1^-}{3}t \right)+\frac{d\rho_1^-}{4}(t-s) \right\}.
\end{eqnarray*}
\end{theorem}
\Proof. Assume $f\in C_\infty^b(M)$ such that $f>0$. Consider $f_n=f h_n$, where $h_n\in C_c^\infty(M)$ is an increasing sequence
with $0< h_n\leq 1$ and $h_n \uparrow 1$ on $M$ as $n\rightarrow\infty$; hence $u_n(t,x):=P_tf_n\uparrow P_tf=:u(t,x)$ for any $(t,x)\in (0,\infty)\times M$.
Since $Lu_n=\partial_t u_n$, by Lemma \ref{gradient}, we have
\begin{eqnarray*}
\Gamma(\log u_n)+\frac{2\rho_2}{3}t\Gamma^Z(\log u_n)\leq \left(\frac{D}{d}+\frac{2\rho_1^-}{3}t\right) \frac{\partial \log u_n}{\partial t}+\frac{d(\rho_1^-)^2}{6}t+\frac{D\rho_1^-}{2}+\frac{D^2}{2dt},
\end{eqnarray*}
or
\begin{equation}\label{harnarck1}
-\left(\frac{D}{d}+\frac{2\rho_1^-}{3}t\right) \frac{\partial \log u_n}{\partial t}\leq -\Gamma(\log u_n)+\frac{d(\rho_1^-)^2}{6}t+\frac{D\rho_1^-}{2}+\frac{D^2}{2dt}.
\end{equation}
Fix $(s,x),(t,y)\in (0,\infty)\times M$ with $s<t$. Let $\gamma: [0,T]\rightarrow M$ be a subunit path such that $\gamma(0)=y$ and $\gamma(T)=x$.
 Consider the path in $(0,\infty)\times M$ defined by
$$\alpha(\tau)=\left(t+\frac{s-t}{T}\tau, \gamma(\tau) \right),\quad 0\leq\tau\leq T;$$
hence $\alpha(0)=(t,y)$ and $\alpha(T)=(s,x)$. By Cauchy-Schwarz inequality, for any $\delta>0$,
\begin{eqnarray*}
&&\log\frac{u_n(s,x)}{u_n(t,y)}=\int_0^T \frac{\d }{\d \tau}\log u_n(\alpha(\tau))\,\d \tau\\
&\leq& \int_0^T \left[\sqrt{\Gamma(\log u_n(\alpha(\tau)))}  -  \frac{t-s}{T}\frac{\partial}{\partial t}\log u_n(\alpha(\tau)) \right]\,\d \tau\\
&\leq& \sqrt{T}\left(\int_0^T \Gamma(\log u_n(\alpha(\tau)))\,\d \tau \right)^{\frac{1}{2}}  -  \frac{t-s}{T}\int_0^T \frac{\partial}{\partial t}\log u_n(\alpha(\tau))\,\d \tau\\
&\leq& \frac{1}{2\delta}T+\frac{\delta}{2}\int_0^T \Gamma(\log u_n(\alpha(\tau)))\,\d \tau - \frac{t-s}{T}\int_0^T \frac{\partial}{\partial t}\log u_n(\alpha(\tau))\,\d \tau.
\end{eqnarray*}
For $\tau \in [0, T]$, let
$$\beta(\tau)=\left(\frac{D}{d}+\frac{2\rho_1^-}{3}\left(t+\frac{t-s}{T}\tau\right)\right).$$
From \eqref{harnarck1}, we obtain
\begin{eqnarray*}
& & - \int_0^T \frac{\partial}{\partial t}\log u_n(\alpha(\tau))\,\d \tau\\
&\leq& - \int_0^T \frac{\Gamma(\log u_n(\alpha(\tau)))}{\beta(\tau)}\,\d \tau + \frac{d(\rho_1^-)^2 }{6}\int_0^T \frac{t+\frac{s-t}{T}\tau}{\beta(\tau)}\,\d \tau\\
& &+\frac{D^2}{2d}\int_0^T \frac{\d \tau}{\beta(\tau)\left(t+\frac{s-t}{T}\tau\right)}+\frac{D\rho_1^- }{2}\int_0^T \frac{\d \tau}{\beta(\tau)}.
\end{eqnarray*}
Choosing $\delta>0$ such that $\frac{\delta}{2}=\frac{t-s}{\beta(0)T}$, we have
\begin{eqnarray*}
\log\frac{u_n(s,x)}{u_n(t,y)}&\leq& \frac{\beta(0)}{4(t-s)}T^2+ \frac{d(\rho_1^-)^2(t-s)}{6T}\int_0^T \frac{t+\frac{s-t}{T}\tau}{\beta(\tau)}\,\d\tau \\
&&+ \frac{D^2(t-s)}{2dT}\int_0^T \frac{\d\tau}{\beta(\tau)\left(t+\frac{s-t}{T}\tau\right)}+ \frac{\rho_1^- D(t-s)}{2T}\int_0^T \frac{\d\tau}{\beta(\tau)}\\
&\leq& \frac{\beta(0)}{4(t-s)}T^2+\frac{d\rho_1^-(t-s)}{4}+\frac{D}{2}\log \left(\frac{t}{s}\right).
\end{eqnarray*}
Minimizing over all subunit paths connecting $y$ to $x$, we get
\begin{eqnarray*}
u_n(s,x)&\leq& u_n(t,y)\left(\frac{t}{s}\right)^{\frac{D}{2}} \exp\left\{\frac{d(x,y)^2}{4(t-s)}\beta(0)+\frac{d\rho_1^-}{4}(t-s) \right\}.
\end{eqnarray*}
Letting $n\rightarrow \infty$, we complete the proof.
\fin

For the heat kernel, we have the following Harnack inequality.
\begin{corollary}\label{kernalharnack}
Assume {\rm(H.1)}, {\rm(H.2)}, {\rm(H.3)} and \eqref{gcdi} holds with $\rho_1<0$. Let $p(t,x,y)$ be the heat kernel on $M$. For any $x,y,z\in M$ and any $0<s<t<\infty$, it holds
\begin{eqnarray*}
 p(s, x, y)&\leq& p(t, x, z)\left(\frac{t}{s}\right)^{\frac{D}{2}} \exp\left\{\left(\frac{D}{d}+\frac{2\rho_1^-}{3}t \right)\frac{d(y,z)^2}{4(t-s)} + \frac{d\rho_1^-}{4}(t-s) \right\}.
\end{eqnarray*}
\end{corollary}
The idea of its proof is to express the heat kernel in terms of the semigroup, i.e.,
$$p(s+\tau, x, y)=P_s(p(\tau,x,\cdot))(y)\,\ \hbox{and}\,\ p(t+\tau, x, z)=P_t(p(\tau,x,\cdot))(z),$$
for $\tau>0$, and then apply Theorem \ref{harnack} (see the proof of Corollary 7.2 in \cite{BaudoinGarofalo}).

From Theorem \ref{harnack}, various mean value type inequalities can be obtained. For example, we have the following one.
\begin{corollary}\label{meanvalueinequality}
Assume {\rm(H.1)}, {\rm(H.2)}, {\rm(H.3)} and \eqref{gcdi} holds with $\rho_1<0$. Given $(s,x), (t,y)$ $\in (0,\infty)\times M$ with $s<t$, for any $f\in C_\infty^b(M)$ such that $f>0$, it holds
\begin{eqnarray*}
(P_sf)(x)&\leq& \frac{1}{\sqrt{\mu\left(B(x,r) \right)}}\left(\int_{B(x,r)} (P_tf)^2(y)\,\d\mu(y) \right)^{\frac{1}{2}}\left(\frac{t}{s}\right)^{\frac{D}{2}}\\ && \times
\exp\left\{\frac{r^2}{4(t-s)}\left(\frac{D}{d}+\frac{2\rho_1^-}{3}t \right)+\frac{d\rho_1^-}{4}(t-s) \right\}.
\end{eqnarray*}
\end{corollary}
\Proof.
By Theorem \ref{harnack}, the Li--Yau type Harnack inequality holds for each $y$ with $d(x,y)=r$. Hence
\begin{eqnarray*}
&&\int_{B(x,r)}(P_sf)^2(x)\,\d\mu(y) \leq \int_{B(x,r)}(P_tf)^2(y)\left(\frac{t}{s}\right)^D \times \\
&&\exp\left\{\frac{2r^2}{4(t-s)}\left(\frac{D}{d}+\frac{2\rho_1^-}{3}t \right)+\frac{2d\rho_1^-}{4}(t-s) \right\}\,\d\mu(y),
\end{eqnarray*}
which implies
\begin{eqnarray*}
&&(P_sf)^2(x)\mu\left(B(x,r) \right)  \leq  \int_{B(x,r)}(P_tf)^2(y)\,\d\mu(y)\left(\frac{t}{s}\right)^D \\ &&\times
\exp\left\{\frac{2r^2}{4(t-s)}\left(\frac{D}{d}+\frac{2\rho_1^-}{3}t \right)+\frac{2d\rho_1^-}{4}(t-s) \right\}.
\end{eqnarray*}
Taking the square root of both sides, we obtain the desired mean value type inequality.\fin

Now we present the main theorem of this section.
\begin{theorem}\label{gaussbound}
Assume {\rm(H.1)}, {\rm(H.2)}, {\rm(H.3)} and \eqref{gcdi} holds with $\rho_1<0$. For any $0<\epsilon<1$, there exist constants $C(d,\kappa,\epsilon,\rho_2)>0$ which tends to $\infty$ as $\epsilon\rightarrow 0^+$,
 and $c(d,\epsilon)>0$, such that for every $x,y\in M$ and $t>0$, it holds
$$  p(t,x,y)\leq\frac{C(d,\kappa,\epsilon,\rho_2)}{\sqrt{\mu\left(B(x,\sqrt{t})\right)\mu\left(B(y,\sqrt{t})\right)}}\exp\left\{c(d,\epsilon)\rho_1^-t - \frac{d(x,y)^2}{(4+\epsilon)t} \right\}.$$
\end{theorem}
For the proof, we follow the method of \cite{CaoYau}, but the idea is originally from \cite{LiYau}.  See also Theorem 7.1 in \cite{BaudoinGarofalo}. For the sake of completeness, we present the proof here.

\Proof. Given $T>0$ and $\delta>0$, we fix $\tau \in (0,(1+\delta)T]$. For a nonnegative function $\psi\in C_c^\infty(M)$ defined in $(0, \tau)\times M$, we consider the function
$$f(t,y)=\int_M p(t,y,z)p(T,x,z)\psi(z)\,\d \mu(z),\quad x\in M.$$
Since $f(t,y)=P_t(p(T,x,\cdot)\psi)(y)$, $y\in M$, it is the solution to the Cauchy problem
\begin{equation*}
\left\{\begin{array}{rl}
Lf-\partial_t f=0,\quad \hbox{in}\,\ (0, \tau)\times M,\\
f(0,z)=p(T,x,z)\psi(z),\quad z\in M.
\end{array}\right.
\end{equation*}
By the hypoellipticity of the operator $L-\partial_t$, $y\mapsto p(T,x,y)$ is $C^\infty(M)$, and hence $p(T,x,\cdot)\psi\in L^\infty(M)$. Moreover, by \eqref{contraction},
\begin{eqnarray*}
\int_M f(t,z)^2\,\d\mu(z)&=&\|P_t(p(T,x,\cdot)\psi)\|^2_{L^2(M)}\leq \|p(T,x,\cdot)\psi\|^2_{L^2(M)}\\
&=&\int_M p(T,x,z)^2\psi(z)^2\,\d\mu(z)<\infty;
\end{eqnarray*}
hence
\begin{eqnarray}\label{gauss1}
\int^\tau_0 \int_M f(t,z)^2\,\d\mu(z)\leq \tau\int_M p(T,x,z)^2\psi(z)^2\,\d\mu(z)<\infty.
\end{eqnarray}
By (4.9) of Corollary 4.6 in \cite{BaudoinGarofalo}, there exists some $\nu\in \R$ such that
$$\Gamma(f)(t,z)\leq e^{-\nu t}\{P_t\Gamma(p(T,x,\cdot)\psi)(z)+P_t\Gamma^Z(p(T,x,\cdot)\psi)(z)\},$$
which implies
\begin{eqnarray}\label{gauss2}
\int^\tau_0 \int_M \Gamma(f)(z,t)\,\d\mu(z)<\infty.
\end{eqnarray}
Now consider a $d$-Lipschitz function $g\in C^1([0,(1+\delta)T)], M)\cap L^\infty([0,(1+\delta)T]\times M)$ such that
\begin{eqnarray}\label{gauss3}
-\frac{\partial g}{\partial t}\geq \frac{1}{2}\Gamma(g),\quad \hbox{on}\,\ [0,(1+\delta)T]\times M.
\end{eqnarray}
Since $f$ is the solution to the above Cauchy problem, we easily get
$$(L-\partial_t)f^2=2f(L-\partial_t)f+2\Gamma(f)=2\Gamma(f).$$
Multiplying this inequality by $h_n^2e^{g}$ with $h_n$ a sequence chosen as in (H.1), we have
\begin{eqnarray*}
0&=&2\int^\tau_0\int_M h_n^2(y)e^{g(t,y)}\Gamma(f)(t,y)\,\d\mu(y)\d t- \int^\tau_0\int_M h_n^2(y)e^{g(t,y)}(L-\partial_t)f(t,y)^2\,\d\mu(y)\d t\\
&=&2\int^\tau_0\int_M h_n^2(y)\left(e^{g}\Gamma(f)\right)(t,y)\,\d\mu(y)\d t + 4\int^\tau_0\int_M h_n^2(y)\left(fe^{g}\Gamma(h_n, f)\right)(t,y)\,\d\mu(y)\d t\\
&+&2\int^\tau_0\int_M h_n^2(y)\left(e^{g}f\Gamma(g, f)\right)(t,y)\,\d\mu(y)\d t - \int^\tau_0\int_M h_n^2(y)\left(e^{g}f^2\frac{\partial g}{\partial t}\right)(t,y)\,\d\mu(y)\d t\\
&-& \int_M h_n^2(y)\left(e^{g}f^2\right)(0,y)\,\d\mu(y) + \int_M h_n^2(y)\left(e^{g}f^2\right)(\tau,y)\,\d\mu(y)\\
&\geq&  2\int^\tau_0\int_M h_n^2e^{g}\left(\Gamma(f)+\frac{1}{4}f^2\Gamma(g)+f\Gamma(f,g)\right)\,\d\mu\d t + 4\int^\tau_0\int_M h_n^2e^{g}\Gamma(h_n,f)\d\mu\d t\\
&-&\int_M h_n^2(y)\left(e^{g}f^2\right)(0,y)\,\d\mu(y) + \int_M h_n^2(y)\left(e^{g}f^2\right)(\tau,y)\,\d\mu(y),
\end{eqnarray*}
where in the second equality we applied integration by parts and in the last inequality we used \eqref{gauss3}. From this we have
\begin{equation}\label{gauss4}
\begin{array}{lll}
\int_M h_n^2(y)\left(e^{g}f^2\right)(\tau,y)\,\d\mu(y)&\leq&\int_M h_n^2(y)\left(e^{g}f^2\right)(0,y)\,\d\mu(y)\\
&&-4\int^\tau_0\int_M h_n^2e^{g}f\Gamma(h_n,f)\d\mu\d t.
\end{array}
\end{equation}
By Cauchy-Schwarz inequality, \eqref{gauss1} and \eqref{gauss2},
\begin{eqnarray*}
&&\left|\int^\tau_0\int_M h_n^2e^{g}f\Gamma(h_n,f)\d\mu\d t\right|\leq \left(\int^\tau_0\int_M h_n^4e^gf^2\Gamma(h_n)\,\d\mu\d t\right)^{\frac{1}{2}}
\left(\int^\tau_0\int_M e^g\Gamma(f)\,\d\mu\d t\right)^{\frac{1}{2}}\\
&&\leq \left(\int^\tau_0\int_M e^gf^2\Gamma(h_n)\,\d\mu\d t\right)^{\frac{1}{2}}\left(\int^\tau_0\int_M e^g\Gamma(f)\,\d\mu\d t\right)^{\frac{1}{2}}\rightarrow 0,
\end{eqnarray*}
as $n\rightarrow \infty$. Hence,
$$\lim_{n\rightarrow \infty}\int^\tau_0\int_M h_n^2e^{g}\Gamma(h_n,f)\d\mu\d t=0.$$
Letting $n\rightarrow \infty$ in \eqref{gauss4}, we obtain
\begin{eqnarray}\label{gauss5}
\int_M e^{g(\tau,y)}f(\tau,y)^2\,\d\mu(y)\leq\int_M e^{g(0,y)}f(0,y)^2\,\d\mu(y).
\end{eqnarray}
Fix $x\in M$ and for $t\in [0,\tau]$. Consider the indicator function $\mathbf{1}_{B(x,\sqrt{t})}$ of the ball $B(x,\sqrt{t})$. Let $\{\psi_k\}_{k\geq 1}$ be a sequence of nonnegative functions in $C_c^\infty(M)$ with $\supp (\psi_k)\subset B(x,90\sqrt{t}) $, such that $\psi_k\rightarrow \mathbf{1}_{B(x,\sqrt{t})}$ as $k\rightarrow \infty$ in $L^2(M)$. For $s\in [0,\tau]$ and any subset $S_1$ of $M$, set
\begin{eqnarray}\label{gauss5+}
f(s,y)=\int_{S_1}p(s,y,z)p(T,x,z)\,\d\mu(z).
\end{eqnarray}
Applying \eqref{gauss5} to $f_k(s,y):=P_s(p(T,x,\cdot)\psi_k)(y)$, we have
\begin{eqnarray}\label{gauss6}
\int_M e^{g(\tau,y)}f_k(\tau,y)^2\,\d\mu(y)  \leq   \int_M e^{g(0,y)}f_k(0,y)^2\,\d\mu(y).
\end{eqnarray}
Since as $k\rightarrow \infty$,
\begin{eqnarray*}
&&\left|\int_M e^{g(\tau,y)}f_k(\tau,y)^2\,\d\mu(y)- \int_M e^{g(\tau,y)}f(\tau,y)^2\,\d\mu(y)  \right|\\
&&\leq 2\|e^{g(\tau,\cdot)}\|_{L^\infty(M)} \|p(T,x,\cdot)\|_{L^2(M)}  \|p(\tau,y,\cdot)\|_{L^\infty(B(x,120\sqrt{t}))}\|\psi_k-\mathbf{1}_{B(x,\sqrt{t})}\|_{L^2(M)}\rightarrow 0,
\end{eqnarray*}
and
\begin{eqnarray*}
&&\left|\int_M e^{g(0,y)}f_k(0,y)^2\,\d\mu(y)  -   \int_M e^{g(0,y)}f(0,y)^2\,\d\mu(y)  \right|\\
&&\leq 2\|e^{g(0,\cdot)}\|_{L^\infty(M)}  \|p(\tau,y,\cdot)\|_{L^\infty(B(x,120\sqrt{t}))}\|\psi_x-\mathbf{1}_{B(x,\sqrt{t})}\|_{L^2(M)}\rightarrow 0,
\end{eqnarray*}
we conclude the same inequality holds with $f_k$ replaced by $f(s,z):=P_s(p(T,x,\cdot)\mathbf{1}_{B(y,\sqrt{t})})(z)$ by letting $k\rightarrow \infty$ in \eqref{gauss6}. It implies the following estimate: for any subset $S_2$ of $M$,
\begin{eqnarray}\label{gauss7}
\begin{array}{lll}
&&\inf_{z\in S_2} e^{g(\tau,z)}\int_{S_2} f(\tau,z)^2\,\d\mu(z)\\
&&\leq \int_{S_2}e^{g(\tau,z)}f(\tau,z)^2\,\d\mu(z) \leq \int_{M}e^{g(\tau,z)}f(\tau,z)^2\,\d\mu(z)\\
&&\leq  \int_{M}e^{g(0,z)}f(0,z)^2\,\d\mu(z) \leq \int_{S_1}e^{g(0,z)}f(0,z)^2\,\d\mu(z)\\
&&\leq \sup_{z\in S_1} e^{g(0,z)}\int_{S_1} f(0,z)^2\,\d\mu(z).
\end{array}
\end{eqnarray}
We choose in \eqref{gauss7} that $S_1=B(y,\sqrt{t})$, $S_2=B(x,\sqrt{t})$ and let
$$g(t,y)=g_x(t,y)=-\frac{d(x,y)^2}{2((1+2\delta)T-t)}.$$
Combining this with the fact that the function $y\mapsto d(x,y)$ satisfies $\Gamma(d(x,\cdot)\leq 1$, we can easily check that the specially chosen $g$ satisfies \eqref{gauss3}. Obviously,
$$\inf_{z\in B(x,\sqrt{t})}e^{g_x(\tau,z)}=\inf_{z\in B(x,\sqrt{t})}\exp\left\{-\frac{d(x,z)^2}{2((1+2\delta)T-\tau)}\right\}\geq \exp\left\{-\frac{t}{2((1+2\delta)T-\tau)}\right\}.$$
Taking $\tau=(1+\delta)T$, from the previous inequality and \eqref{gauss5+}, we obtain
\begin{equation}\label{gauss8}
\begin{array}{lll}
&&\int_{B(x,\sqrt{t})}f((1+\delta)T,z)^2\,\d\mu(z)\\
&\leq& \left(\sup_{z\in B(y,\sqrt{t})}\exp\left\{ -\frac{d(x,z)^2}{2(1+2\delta)T}+\frac{t}{2\delta T} \right\}\right)\int_{B(y,\sqrt{t})}p(T,x,z)^2\,\d\mu(z).
\end{array}
\end{equation}
Applying Theorem \ref{harnack}, for every $z\in B(x,\sqrt{t})$,
$$f(T,x)^2\leq f((1+\delta) T),z)^2(1+\delta)^D\exp\left\{\frac{t}{2\delta T}\left(\frac{D}{d}+\frac{2\rho_1^-}{3}(1+\delta)T \right) + \frac{\delta d\rho_1^-T}{2}  \right\}.$$
Integrating on both sides in $z$ on $B(x,\sqrt{t})$, we have
\begin{equation*}
\begin{array}{lll}
&&\left(\int_{B(y,\sqrt{t})} p(T,x,z)^2\,\d\mu(z)\right)^2 = f(T,x)^2\\
&\leq&\frac{(1+\delta)^D\exp\left\{\frac{t}{2\delta T}\left(\frac{D}{d}+\frac{2\rho_1^-}{3}(1+\delta)T\right) + \frac{\delta d\rho_1^- T}{2}   \right\}}{\mu\left(B(x,\sqrt{t})\right)}\int_{B(x,\sqrt{t})}f((1+\delta)T,z)^2\,\d\mu(z)\\
&\leq& \frac{(1+\delta)^D\exp\left\{\frac{t}{2\delta T}\left(\frac{D}{d}+\frac{2\rho_1^-}{3}(1+\delta)T\right) + \frac{\delta d\rho_1^- T}{2}  \right\}}{\mu\left(B(x,\sqrt{t})\right)}\sup_{z\in B(y,\sqrt{t})}\exp\left\{ -\frac{d(x,z)^2}{2(1+2\delta)T}+\frac{t}{2\delta T} \right\}\\
&&\times\int_{B(y,\sqrt{t})}p(T,x,z)^2\,\d\mu(z),
\end{array}
\end{equation*}
where we used \eqref{gauss8} in the second inequality.
Hence,
\begin{eqnarray*}
\begin{array}{lll}
\int_{B(y,\sqrt{t})} p(T,x,z)^2\,\d\mu(z)&\leq& \frac{(1+\delta)^D\exp\left\{\frac{t}{2\delta T}\left(\frac{D}{d}+\frac{2\rho_1^-}{3}(1+\delta)T\right) + \frac{\delta d\rho_1^- T}{2}   \right\}}{\mu\left(B(x,\sqrt{t})\right)}\\
&&\times\sup_{z\in B(y,\sqrt{t})}\exp\left\{ -\frac{d(x,z)^2}{2(1+2\delta)T}+\frac{t}{2\delta T} \right\}.
\end{array}
\end{eqnarray*}
Choosing $T=(1+\delta)t$ in the above inequality, we get
\begin{eqnarray}\label{gauss9}
\begin{array}{lll}
\int_{B(y,\sqrt{t})} p((1+\delta)t,x,z)^2\,\d\mu(z)&\leq& \frac{(1+\delta)^D\exp\left\{\frac{1}{2\delta (1+\delta)t}\left(\frac{D}{d}+\frac{2\rho_1^-}{3}(1+\delta)^2t\right) + \frac{\delta(1+\delta)d\rho_1^- t}{2}   \right\}}{\mu\left(B(x,\sqrt{t})\right)}\\
&&\times\sup_{z\in B(y,\sqrt{t})}\exp\left\{ -\frac{d(x,z)^2}{2(1+2\delta)(1+\delta)t}+\frac{1}{2\delta (1+\delta)} \right\}.
\end{array}
\end{eqnarray}
Applying Corollary \ref{kernalharnack}, we obtain, for every $z\in B(y,\sqrt{t})$,
$$p(t,x,y)^2\leq p((1+\delta)t,x,z)^2(1+\delta)^D\exp\left\{\frac{1}{2\delta}\left(\frac{D}{d}+\frac{2\rho_1^-}{3}(1+\delta)t\right) + \frac{\delta d\rho_1^- t}{2}   \right\}.$$
Integrating this inequality in $z$ on $B(y,\sqrt{t})$, we have
\begin{eqnarray*}
\mu\left(B(y,\sqrt{t})\right)p(t,x,y)^2&\leq& (1+\delta)^D\exp\left\{\frac{1}{2\delta}\left(\frac{D}{d}+\frac{2\rho_1^-}{3}(1+\delta)t\right) + \frac{\delta d\rho_1^- t}{2}   \right\}\\
&&\times\int_{B(y,\sqrt{t})}p((1+\delta)t,x,z)^2\,\d\mu(z).
\end{eqnarray*}
Combining this inequality with \eqref{gauss9}, we get
\begin{eqnarray*}
p(t,x,y)&\leq& \frac{(1+\delta)^D\exp\left\{\frac{2+\delta}{4\delta(1+\delta)}\frac{D}{d} + \frac{\rho_1^-(1+\delta)t}{3\delta} + \frac{\delta(2+\delta) d\rho_1^- t}{4} + \frac{1}{4\delta(1+\delta)}   \right\}}{\sqrt{\mu\left(B(x,\sqrt{t})\right)\mu\left(B(y,\sqrt{t})\right)}}\\
&& \times   \sup_{z\in B(y,\sqrt{t})}\exp\left\{-\frac{d(x,z)^2}{4(1+2\delta)(1+\delta)t}\right\}.
\end{eqnarray*}
If now $x\in  B(y,\sqrt{t})$, then for $z\in  B(y,\sqrt{t})$, $d(x,z)^2>d(x,y)^2-t$, and hence
$$\sup_{z\in B(y,\sqrt{t})}\exp\left\{-\frac{d(x,z)^2}{4(1+2\delta)(1+\delta)t}\right\}\leq \exp\left\{-\frac{d(x,y)^2}{4(1+2\delta)(1+\delta)t}+\frac{1}{4(1+2\delta)(1+\delta)}\right\}.$$
If $x\notin  B(y,\sqrt{t})$, then $d(x,y)\geq \sqrt{t}$. Hence, $d(x,z)^2\geq \left(d(x,y)-\sqrt{t}\right)^2\geq \frac{d(x,y)^2}{1+\delta}-\frac{t}{\delta}$. Thus,
$$\sup_{z\in B(y,\sqrt{t})}\exp\left\{-\frac{d(x,z)^2}{4(1+2\delta)(1+\delta)t}\right\}\leq \exp\left\{-\frac{d(x,y)^2}{4(1+2\delta)(1+\delta)^2t}+\frac{2+\delta^{-1}}{4(1+2\delta)(1+\delta)}\right\}.$$
For any $\epsilon>0$, we choose $\delta>0$ such that $4(1+2\delta)(1+\delta)^2=4+\epsilon$. Therefore, we complete the proof. \fin

\section{Upper bounds for $p(t,x,y)$ under $CD(\rho_1,\rho_2,\kappa, \infty)$}
In this section, we aim to develop another argument to establish the upper bound for the heat kernel $p(t,x,y)$ under the generalized curvature inequality \eqref{gci}. To be precise, we unify Grigor'yan's integrated maximum principle and Wang's dimension-free Harnack inequality to achieve our goal.

The following key inequality in this section, also called integrated maximum principle was proved by Grigor'yan \cite{Grigor'yan} for the Laplacian case on the Riemannian manifolds, i.e., $L=\Delta$. We prove its generalization in the sub-Riemannian setting here.
\begin{lemma}\label{grigoyan}
For $x\in M$, $T>0$, $p>1$ and $q=\frac{p}{2(p-1)}$, let
$$\eta(s,y)=-\frac{d(x,y)^2}{2(T-qs)},\quad y\in M,\,\ s<\frac{T}{q}.$$
Then, for any $f\in L^\infty(M)$ with $f\geq 0$,
$$\int_M \left(P_tf\right)^p(y)e^{\eta(t,y)}\,\d\mu(y)\leq \int_M f(y)^pe^{-\frac{d(x,y)^2}{2T}}\,\d\mu(y),\quad 0\leq t<\frac{T}{q}.$$
\end{lemma}
\Proof. In view of Section 4 in \cite{GrigoryanHu}, it suffices to prove that, for any relatively compact open subset $\Omega$ of $M$,
$$\int_\Omega \left(P_t^\Omega f\right)^p(y)e^{\eta(t,y)}\,\d\mu(y)\leq \int_\Omega f(y)^pe^{-\frac{d(x,y)^2}{2T}}\,\d\mu(y),\quad 0\leq t<\frac{T}{q},$$
where $P_t^\Omega$ is the semigroup of the Dirichlet form of $L$ restricted to $\Omega$ (see \cite{FukushimaOshima}).
Let
$$I_\Omega(s)=\int_\Omega \left(P_s^\Omega f(y)\right)^pe^{\eta(s,y)}\,\d\mu(y),\quad 0\leq s< \frac{T}{q}.$$
Since $u(s,\cdot):=P^\Omega_s f\in L^2(\Omega)$ and $\eta(s,\cdot)$ is bounded in $\Omega$, the function $I_\Omega(s)$ is finite.
Notice that $I_\Omega(s)$ is continuous in $s\in [0,\frac{T}{q})$ in $L^2(\Omega)$ and the function $s\mapsto e^{\eta(s,\cdot)}$ is
obviously continuous in $s$ in the sup norm in $C_b(\Omega)$, from which the function $s\mapsto u(s,\cdot)e^{\eta(s,\cdot)}$ is continuous in $s\in [0,\frac{T}{q})$ in $L^2(\Omega)$.

Now it is suffices to prove that $I^{'}_\Omega(s)$ exists and is non-positive for all $s\in (0,\frac{T}{q})$. Fix some $s\in (0,\frac{T}{q})$. Since $\eta(s,\cdot)$, $\frac{\partial\eta}{\partial s}(s,\cdot)$ are continuous and bounded in $\bar{\Omega}$, the complement of $\Omega$, they both belong to $C_b(\Omega)$. Hence, $\frac{\partial\eta}{\partial s}$ is the strong derivative in $C_b(\Omega)$ and so is $e^{\eta(s,\cdot)}$. As we know, $u(s,\cdot)$ is strongly differentiable in $L^2(\Omega)$ and its strong derivative $\frac{\partial u}{\partial s}$ in $L^2(\Omega)$ is given by $\frac{\partial u}{\partial s} =L u$. Hence, $u(s,\cdot)e^{\eta(s,\cdot)}$ is strongly differentiable in $L^2(\Omega)$. Thus, for $0< s< \frac{T}{q}$,
$$I^{'}_\Omega(s)=\int_\Omega p\left(u(s,y)\right)^{p-1} L u(s,y)e^{\eta(s,y)}\,\d\mu(y) - \int_\Omega\left(u(s,y)\right)^p e^{\eta(s,y)}\frac{qd(x,y)^2}{2(T-qs)^2}\,\d\mu(y).$$
By the chain rule for Lipschitz functions, $e^{\eta(s,\cdot)}$ is a Lipschitz function. Since $e^{\eta(s,\cdot)}$ is bounded and Lipschitz in $\Omega$ and $u(s,\cdot)$ belongs to the Sobolev space $W_{0}^{1,2}(\Omega)$ (see \cite{GarofaloNhieu} for the definition), we have $u(s,\cdot)e^{\eta(s,\cdot)}\in W_{0}^{1,2}(\Omega)$. Hence, the first term on the right hand side of the last inequality equals
$$-p\int_\Omega \Gamma(u^{p-1}e^\eta,u)\,\d\mu.$$
By the fact that $\Gamma(d(x,\cdot))\leq 1$ $\mu$-a.e. for any fixed $x\in M$. Combining with the Cauchy-Schwarz inequality, we have
\begin{eqnarray*}
\int_\Omega \Gamma(u^{p-1}e^\eta,u)\,\d\mu&=& \int_\Omega \left[ -u^{p-1}e^\eta\frac{d}{T-sq}\Gamma(d,u)+(p-1)u^{p-2}e^{\eta}\Gamma(u)\right]\,\d\mu\\
&\geq& \int_\Omega \left[-\frac{u^{p-1}e^\eta d}{T-sq}\Gamma(u)^{\frac{1}{2}}+(p-1)u^{p-2}e^{\eta}\Gamma(u)\right]\,\d\mu,
\end{eqnarray*}
where we write the distance function as $d$ for short. Thus, since $q=\frac{p}{2(p-1)}$, we have
$$I^{'}_\Omega(s)\leq -p(p-1)\int_\Omega (P_s^\Omega f)^pe^{\eta(s,\cdot)}\left[\frac{d(x,\cdot)}{2(p-1)(T-sq)^2}-\frac{\Gamma(P_s^\Omega f)^{\frac{1}{2}}}{P_s^\Omega f}\right]^2\,\d\mu\leq 0.$$
Therefore, we complete the proof by integrating the above inequality over $[0,t]$.\fin

The dimension-free Harnack inequality is introduced by F.-Y. Wang in \cite{Wany1} in the Riemannian setting under the assumption of curvature inequality, which is equivalent to the assumption of a lower bound on the Ricci curvature. See also \cite{ArnaudonThalmaierWang,Wany2,Wang2+,Wang2++,Wang2+++,WanyYuan} for example and references therein for extensive study on this inequality and its various applications.  Recently, Wang's Harnack inequality is established under the generalized curvature inequality in the sub-Riemannian setting in \cite{BaudoinBonnefont1}. We present it here as a proposition.
\begin{proposition}\label{wanyharnack}
Assume {\rm(H.1)}, {\rm(H.2)}, {\rm(H.3)} and \eqref{gci} holds with $\rho_1\in \R$. For any $f\in L^\infty(M)$ with $f\geq 0$, $p>1$, $t>0$ and $x,y\in M$,
$$(P_tf)^p(x)\leq P_tf^p(y)\exp\left\{\frac{p}{p-1}\left(\frac{1+\frac{2\kappa}{\rho_2}+2\rho^-_1 t}{4t}\right) d(x,y)^2 \right\}.$$
\end{proposition}

Now we presented the main result of this section in the following theorem.
\begin{theorem}\label{freebound}
Assume {\rm(H.1)}, {\rm(H.2)}, {\rm(H.3)} and \eqref{gci} holds with $\rho_1\in \R$. For any $\sigma>2$, there is a constant $c(\sigma)>0$ such that
$$p(t,x,y)\leq \frac{1}{\sqrt{\mu\left(B(x,\sqrt{t})\right)\mu\left(B(y,\sqrt{t})\right)}}\exp\left\{c(\sigma)(1+t)-\frac{d(x,y)^2}{2\sigma t}\right\}.$$
\end{theorem}
The proof is based on the integrated maximum principle in Lemma \ref{grigoyan} and the dimension-free Harnack inequality in Proposition \ref{wanyharnack} (see e.g. \cite{GongWang} for the Riemanian case).

\Proof. For $\sigma>2$, let $p\in (1,2)$ such that $q:=\frac{p}{2(p-1)}<\frac{\sigma}{2}$, and let $T=\frac{\sigma t}{2}$. Applying Proposition \ref{wanyharnack}, for $f\in L^\infty(M)$ with $f\geq 0$,
\begin{eqnarray*}
&&\mu\left(B(x,\sqrt{2t})\right)(P_tf)^2(x) \exp\left\{-\frac{p^2}{2-p}\left(1+\frac{2\kappa}{\rho_2}+2\rho_1^- t  \right)-\frac{1}{\frac{\sigma}{2}-q}   \right\}\\
&\leq& \int_M (P_tf)^2(x)\exp\left\{-\frac{2p^2}{2-p}\left(\frac{1+\frac{2\kappa}{\rho_2}+2\rho^-_1 t}{4t}\right) d(x,y)^2- \frac{d(x,y)^2}{2(T-qt)}  \right\}\,\d\mu(y)\\
&\leq& \int_M \left(P_tf^{\frac{2}{p}}\right)^p(y)\exp\left\{- \frac{d(x,y)^2}{2(T-qt)}  \right\}\,\d\mu(y)\leq \int_M f^2(y)e^{-\frac{d(x,y)^2}{2T}}\,\d\mu(y),
\end{eqnarray*}
where we applied Lemma \ref{grigoyan} in the last inequality. For $n\geq 1$, take
$$f(y)=\left(n\wedge p(t,x,y) \right)e^{\frac{n\wedge d(x,y)^2}{2T}},\quad y\in M.$$
Then
$$(P_tf)^2(x)=\left(\int_M p(t,x,y)f(y)\,\d\mu(y) \right)^2\geq \left(\int_M \left(n\wedge p(t,x,y)\right)^2 e^{\frac{n\wedge d(x,y)^2}{2T}}\,\d\mu(y) \right)^2,$$
and
$$\int_M f^2(y)e^{-\frac{d(x,y)^2}{2T}}\,\d\mu(y)\leq \int_M \left(n\wedge p(t,x,y)\right)^2 e^{\frac{n\wedge d(x,y)^2}{2T}}\,\d\mu(y).$$
Hence,
$$\int_M \left(n\wedge p(t,x,y)\right)^2 e^{\frac{n\wedge d(x,y)^2}{\sigma t}}\,\d\mu(y)\leq \frac{e^{c(\sigma)(1+2t)}}{\mu\left(B(x,\sqrt{2t})\right)},$$
for some constant $c(\sigma)>0$. Letting $n\rightarrow \infty$, we arrive at
$$\int_M  p(t,x,y)^2 e^{\frac{d(x,y)^2}{\sigma t}}\,\d\mu(y)\leq \frac{e^{c(\sigma)(1+2t)}}{\mu\left(B(x,\sqrt{2t})\right)}.$$
Applying the last inequality for $\frac{t}{2}$ in place of $t$, we get
\begin{eqnarray*}
&&p(t,x,y)=\int_M p\left(\frac{t}{2},x,z\right) p\left(\frac{t}{2},y,z\right)\,\d\mu(z)\\
&\leq& e^{-\frac{d(x,y)^2}{2\sigma t}}\int_M p\left(\frac{t}{2},x,z\right)e^{\frac{d(x,z)^2}{2\sigma t}}   p\left(\frac{t}{2},y,z\right)e^{\frac{d(y,z)^2}{2\sigma t}}\,\d\mu(z)\\
&\leq& e^{-\frac{d(x,y)^2}{2\sigma t}}  \left(\int_M p\left(\frac{t}{2},x,z\right)e^{\frac{d(x,z)^2}{2\sigma t}}\,\d\mu(z)\right)^{\frac{1}{2}}  \left(\int_M p\left(\frac{t}{2},y,z\right)e^{\frac{d(y,z)^2}{2\sigma t}}\,\d\mu(z)\right)^{\frac{1}{2}}\\
&\leq& \frac{1}{\sqrt{\mu\left(B(x,\sqrt{t})\right)\mu\left(B(y,\sqrt{t})\right)}}\exp\left\{c(\sigma)(1+t)-\frac{d(x,y)^2}{2\sigma t}\right\}.
\end{eqnarray*}
Therefore, we complete the proof.\fin

We remark here that the upper bound in Theorem \ref{freebound} is independent of the dimension.

\end{document}